# Towards a copilot in BIM authoring tool using a large language model-based agent for intelligent human-machine interaction

Changyu Du, Stavros Nousias, André Borrmann
Technical University of Munich, Germany
changyu.du@tum.de

**Abstract.** Facing increasingly complex BIM authoring software and the accompanying expensive learning costs, designers often seek to interact with the software in a more intelligent and lightweight manner. They aim to automate modeling workflows, avoiding obstacles and difficulties caused by software usage, thereby focusing on the design process itself. To address this issue, we proposed an LLM-based autonomous agent framework that can function as a copilot in the BIM authoring tool, answering software usage questions, understanding the user's design intentions from natural language, and autonomously executing modeling tasks by invoking the appropriate tools. In a case study based on the BIM authoring software Vectorworks, we implemented a software prototype to integrate the proposed framework seamlessly into the BIM authoring scenario. We evaluated the planning and reasoning capabilities of different LLMs within this framework when faced with complex instructions. Our work demonstrates the significant potential of LLM-based agents in design automation and intelligent interaction.

## 1. Introduction

Modern BIM authoring software has become increasingly complex and heavy due to its ability to cover design requirements in various disciplines. This complexity requires designers to undergo extensive training and gain experience to master and understand software operations, enabling them to translate their design intentions into a stream of commands within the software. This significantly raises the bar for using the software, creating obstacles in adopting BIM-based model design.

An autonomous agent is a system that functions within a certain environment. It senses the environment around it and aims to accomplish tasks through self-directed planning and actions (Franklin and Graesser, 1997). Such agent systems are often found in reinforcement learning scenarios, where the agent acts according to simple heuristic policy functions and learns in isolated and restricted environments, which makes it often difficult for the agent to replicate human-level decision-making in open environments (Wang *et al.*, 2024). Recently, due to the remarkable natural language understanding and almost human-like intelligence demonstrated by Large Language Models (LLMs), increasing numbers of research in both industry and academia have focused on LLM-based autonomous agents. The core idea is to equip LLMs with tools that enable interaction with the external world, allowing them to behave, plan and complete tasks like humans. Microsoft 365 Copilot[1] works within Microsoft Office apps like Word, Excel and Outlook, assisting in tasks such as drafting documents, summarizing emails and plotting tables by leveraging LLMs and data context. GitHub Copilot[2] assists programmers by automatically completing comments and writing code. Through a chat interface embedded in the IDE, developers can interact with it to analyze and explain the purpose of code blocks, generate unit tests, and even receive suggestions for fixing errors. Such application of LLM-based agents in various fields has significantly improved user efficiency, sparking our interest in researching their use as design assistants in BIM authoring scenarios.

---

[1] https://blogs.microsoft.com/blog/2023/03/16/introducing-microsoft-365-copilot-your-copilot-for-work/
[2] https://github.blog/2023-03-22-github-copilot-x-the-ai-powered-developer-experience/

To liberate designers from the additional effort of translating design ideas into software commands and make the interaction with BIM design software more intelligent, we propose an autonomous agent framework based on LLMs. The agent can infer user complex intents from natural language and autonomously execute appropriate workflows in BIM design software by interacting with the underlying APIs, as well as answer software usage questions by accessing external knowledge bases. In a case study, we demonstrate the seamless integration of the proposed agent into the BIM authoring software Vectorworks, and conduct detailed experiments to comprehensively evaluate the performance of different LLMs under this framework for various tasks. Our research proposes a new way of interacting with design software and lays the groundwork for implementing intelligent copilots in BIM authoring tools.

## 2. Related works

Recent advancements in large language models (LLMs) have shown remarkable potential in mimicking human intelligence. This success is primarily due to extensive training datasets and a large number of model parameters. There is increasing research in using LLMs as the main component in autonomous agents, aiming to achieve human-like decision-making. Researchers are focusing on integrating human-like capabilities such as memory and planning into LLMs, enabling them to perform a range of tasks efficiently (Wang *et al.*, 2024). HuggingGPT (Shen *et al.*, 2023) proposed a framework that leverages LLMs to connect various AI models in machine learning communities to solve AI tasks. 3D-GPT (Sun *et al.*, 2023) is a framework that leverages LLMs for 3D modeling based on instructions, employing multiple agents to process and execute tasks collaboratively. Their work enhances initial scene descriptions into detailed forms and uses code generation to interface with 3D software like Blender, facilitating asset creation. Mehta et al. introduced an interactive framework that enables human architects and agents to collaboratively construct structures in a 3D simulation environment similar to Minecraft (Mehta *et al.*, 2024). The interactive agents can comprehend natural language instructions, place blocks, seek clarifications, and incorporate human feedback.

A recent study (Jang *et al.*, 2023) introduced using an LLM-based design assistant to detail the exterior walls in Revit automatically. Their work builds upon their previous research (Jang and Lee, 2022), which essentially involves converting BIM models into a textual representation in XML/JSON format, then using LLMs to modify architectural details within the structured text, and finally converting the modified text back into BIM models. In their latest study, an interactive Revit plugin was developed to leverage LLMs and prompt engineering to extract relevant information from dialogues for subsequent structured text processing. Experiments were designed to analyze whether the generated wall details meet the designers' requirements and thermal engineering standards. In contrast, the method framework we propose is applicable to more general scenarios and can be extended to a broader range of use cases.

## 3. Methodology

Our work was inspired by the HuggingFace Transformer Agent[3]. The main idea is to utilize prompt engineering techniques to enable state-of-the-art LLMs such as GPT-4 (OpenAI, 2023) and Mixtral-8×7B (Jiang *et al.*, 2024) with strong in-context learning capabilities to generate Python code that can interact with BIM authoring software. Our proposed method does not require retraining or fine-tuning the LLM, thus saving on expensive computational costs.

---

[3] https://huggingface.co/docs/transformers/transformers_agents

The overall pipeline with a sample input is shown in Figure 1. The input instruction can be in text or voice form. Incorporating the speech-to-text model Whisper (Radford *et al.*, 2023) into our workflow, the voice instruction is automatically converted into text for input into the subsequent prompt template. The prompt template is designed to enable the LLM agent to invoke suitable predefined tool functions in the generated Python code to complete tasks specified by humans. In the predefined tool set, we provided various tool functions (create a wall, move, delete, etc. ) and textual descriptions of their functionalities. These tools encapsulate the underlying APIs of BIM software and cover different interaction types, complexities, and capabilities. Our framework does not restrict the choice of LLMs; theoretically, any LLMs capable of generating Python code can be supported. Based on the completed prompt template, the LLM agent will provide reasoning and write the appropriate code, which is then evaluated and executed by a custom Python interpreter with syntax checking. Finally, the returned result is displayed in the BIM software. We also implemented a memory module to store past chat histories, code execution results, and defined variables. This allows the agent to have comprehensive contextual information and feedback from the environment during sessions, enabling it to refine and improve its responses in conversations with humans.

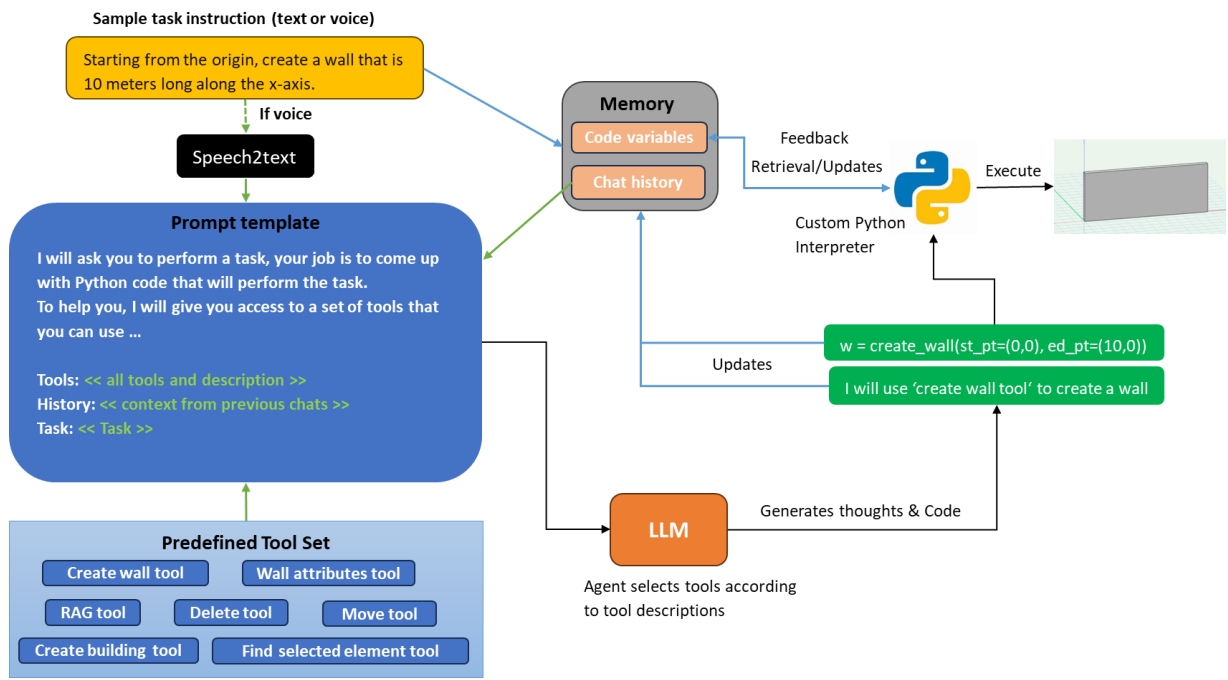

Figure 1 The proposed LLM-based tool agent pipeline with an example task

## 3.1 Prompt template

The details of the prompt template used in this study are shown in Figure 2. We assigned a role to the LLM and defined specific requirements to guide the LLM in responding to user instructions as expected. In this template, we employed prompt engineering techniques such as chain-of-thought (Wei *et al.*, 2024), which involves making the model infer step by step and use intermediate steps to achieve better complex reasoning capabilities. Additionally, we used zero-shot prompting (Wei *et al.*, 2021), leveraging the characteristic of current LLMs being instruction-tuned on vast amounts of code data, allowing them complete code generation tasks directly based on the given prompt description without providing extra examples in the prompt template, known as few-shot prompting techniques (Brown *et al.*, 2020). We experimented with few-shot prompts but did not observe a significant difference. We believe that in our scenario, zero-shot prompting can save tokens amount while ensuring high-quality responses.

> You are an AI assistant that specializing in coding in Vectorworks. You try to be helpful, polite, honest, and humble-but-knowledgeable.
> The job of the AI assistant is to write Python code that invoke suitable pre-defined tool functions to complete tasks specified by humans in Vectorworks. Each instruction in Python should ideally be a simple assignment.
> To help with that, you have access to a set of tools. Each tool is a Python function and has a description explaining the task it performs, the inputs it expects and the outputs it returns.
> You should first think step-by-step - describe your plan for which tools you will use to perform the task and why. Then output the Python code in a single code block. Minimize any other prose. You can only give one reply for each conversation turn.
>
> Tools:
> <<tool function names and descriptions>>
> In your code, you should only use the tools mentioned above that are accessible to you, rather than attempting to invent new tools. Avoid making any assumptions in code. If necessary, you can import and use the Python standard library.
>
> Now, based on the conversation given below, please complete your answer as an AI assistant:
> <<chat history>>
> Human: <<task>>
> Assistant:

Figure 2 Prompt template. Placeholders are marked with colors, indicating the dynamic content that can be inserted into the template.

## 3.2 Toolset

Table 1 presents predefined tool functions in the toolset and their descriptions. We designed various representative tools for agents from three perspectives, essentially covering basic aspects of human-machine interaction in BIM authoring software: **a)** CRUD operations for building components (marked in blue), **b)** Creating complex models with parametric tools (marked in green), **c)** Addressing software usage questions based on external documentation (marked in red).

Table 1 Implemented tools in this study

| Tool function name | Description |
|---|---|
| Create_wall | This tool is used to create a wall in Vectorworks. It takes two inputs: 'st_pt', which should be the start point of the wall, and 'ed_pt', which should be the end point of the wall. Both 'st_pt' and 'ed_pt' are 2D coordinates string. This tool will return nothing. |
| Set_wall_attributes | This tool is used to set the geometric properties of a wall, such as thickness, height and offset. The required input is a wall's uuid. Depending on the specific property settings needed, optional input parameters include 'thickness', 'height', and 'bot_offset'. The 'bot_offset' refers to the vertical distance from the bottom of the wall to the XY plane at the origin. This tool will return the uuid of the modified wall object. |
| Move | This tool is used to move a list of elements in Vectorworks. It takes four required inputs: the 'xDistance', 'yDistance', 'zDistance', and 'uuid'. These represent moving distance in x, y, z directions, and the element's unique uuid. The moving distances in each direction should be either integer or float values. The 'uuid' can be a list or a single string. |
| Delete | This tool deletes an element or a list of elements in Vectorworks. It takes an element's unique uuid or a list of uuids as input and then deletes the elements. |
| Find_selected_element | This tool is used to get selected elements in Vectorworks. It takes no input but returns the elements uuids in the list. If no elements are found, it will return an empty list. |
| Create_building | This tool is used to create a building from the selected floorplan shape. As input, it takes a single polygon shape's uuid and the styles of wall, slab and roof slab. Also, the user can specify the story height and story amount. The complete list of input parameters and their corresponding types are: "floorplan_shape_uuid: str, slab_style_name_roof: str, slab_style_name: str, wall_style_name_first_floor: str, wall_style_name: str, number_of_stories: int, wall_height_first_floor: int, |
| | story_height: int." Only the floorplan_shape_uuid is the required parameter, and the rest are optional. If not specified, the default values will be used. The return value is the height of the building. |
| Document_retrieval | This tool is designed to aid the assistant in responding to user inquiries specifically related to software usage questions about Vectorworks. When the assistant identifies a question that requires detailed, accurate information from the official documentation of Vectorworks, this tool should be invoked to search for and provide the necessary information to the user. The tool takes one input string: 'question', which should be the user's question. Formulate the input to the tool using the exact phrasing of the user's question whenever possible. Ensure to maintain the context and specificity the user provided to retrieve the most relevant section of the documentation. |

Each blue tool invokes the relevant Python APIs of Vectorworks based on its design requirements. As the raw APIs are often fine-grained and low-level, each tool intrinsically encapsulates the logic that combines different APIs to achieve the tool's functionality. The green tool is essentially an encapsulation of a script for generating parametric buildings. We initially developed a parametric building object using the Marionette graphical scripting tool in Vectorworks (similar to Dynamo/Grasshopper), which allows for the generation of different design variations by changing parameters such as the floor plan shape, number of stories, wall styles, etc. We then exported it as a Python script and wrapped it in a Python function, aligning the function and building parameters.

The red-labeled document retrieval tool is designed to answer user inquiries regarding software usage. To prevent LLMs from generating unreliable responses due to hallucinations, we expect LLMs to refer to Vectorworks' documentation when answering software usage questions. This tool, therefore, effectively encapsulates a Retrieval Augmentation Generation (RAG) (Lewis *et al.*, 2020) workflow, a technique that enhances the accuracy and reliability of LLMs by leveraging facts obtained from external knowledge sources. We can extract relevant snippets from Vectorworks documentation through RAG and enable the LLM to generate dependable answers based on this content.

The custom RAG workflow behind the tool is shown in Figure 3. We first collected and cleaned 1911 HTML files of Vectorworks online documentation[4], removing unnecessary hyperlinks, images, etc., and converted them into Markdown format. This allowed the documents' structure and tables to be represented in plain text using Markdown syntax. Given that each Markdown document is generally not very long and contains numerous tables, we did not further split the document into smaller chunks in order to retain table structure and contextual coherence. The processed documents are fed into a pre-trained Sentence Transformer model (Reimers and Gurevych, 2019) to obtain their corresponding embedding vector representations. This process is crucial as it converts text into a numerical format that machines can understand, allowing for more nuanced and complex interpretations of language beyond simple keyword matching. These vectors, representing the essence of the documents in high-dimensional space, are indexed and stored in a vector database for later retrieval. Given a user query, the embedding model first converts it into a vector representation. This step is essential for aligning the user's request with the same numerical space as the documents. We then search for the two nearest neighbors of this query vector in the vector database by calculating the cosine similarity between vectors, which in practice means finding documents with the most similar content to the query. The Markdown text of these two candidates, along with the user query text, is then input into the prompt. GPT-4 is ultimately asked to answer the question based on the given context from the most relevant documents, ensuring the accuracy and reliability of the response. It's worth noting that any other LLM can easily replace GPT-4 in this workflow.

---

[4] https://app-help.vectorworks.net/2023/eng/VW2023_Guide/LandingPage/Welcome_to_Vectorworks.htm

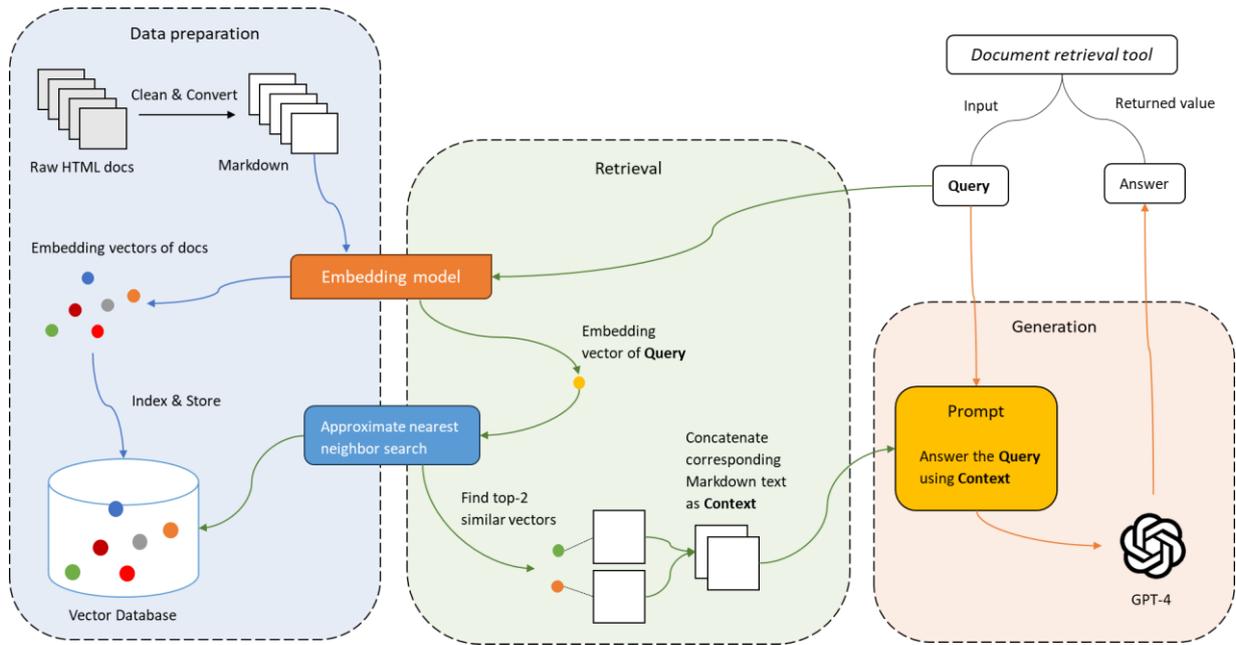

Figure 3 The implemented RAG workflow behind the document retrieval tool. "Query" is the input parameter of the tool function, which is a question from the user about the software usage, and "Answer" is the return value of the tool, which is a textual answer to the software usage issue generated by the LLM based on the retrieved context. Please note that here, GPT-4 and the prompt operate independently of the agent framework and are solely dedicated to this workflow. In addition, GPT-4 can be replaced by any other LLMs.

### 3.3 Custom interpreter

The custom Python interpreter is designed to evaluate Python code in a controlled environment. It uses the ast (Abstract Syntax Tree) module to parse code into a tree of nodes and then evaluate it. A state dictionary stores and tracks defined variables and their values during the evaluation. This information will also be stored in the memory module, allowing LLMs to access variables defined in previous code within the current conversation round, thereby maintaining a comprehensive context throughout the session. We significantly expanded the limited Python interpreter from the Transformer Agent framework, ensuring it supports standard Python features while safely executing code. For instance, it restricts importing arbitrary third-party libraries, only allowing the Python standard library and predefined tool functions. Additionally, it limits file I/O, multithreading, and network operations, preventing the execution of potentially harmful code in the system.

### 4. Case study

Our case study features the BIM authoring tool Vectorworks, where we integrated the proposed framework by developing a web palette plugin using the architecture shown in Figure 4. The C++ backend of the web palette allows defining JavaScript functions, enabling the frontend implemented by Vue.js to call them. This allows the implementation of a dynamic web interface embedded within Vectorworks. Since our framework is entirely based on Python, we invoke the built-in Python engine of Vectorworks on the backend to execute our code, thus delegating the JavaScript implementation. We utilize the memory module to store the state between Python calls. Figure 5 shows the developed prototype. The user can directly chat with the agent by clicking the microphone button, and the backend automatically calls the Whisper model (Radford *et al.*, 2023) to convert the audio to text and populate the input message box. This adds a new dimension to interaction with the BIM authoring tool. The LLMs we used in this study are the latest version of GPT-4 (OpenAI, 2023) and the open-soure model Mixtral-8×7B (Jiang *et al.*, 2024).

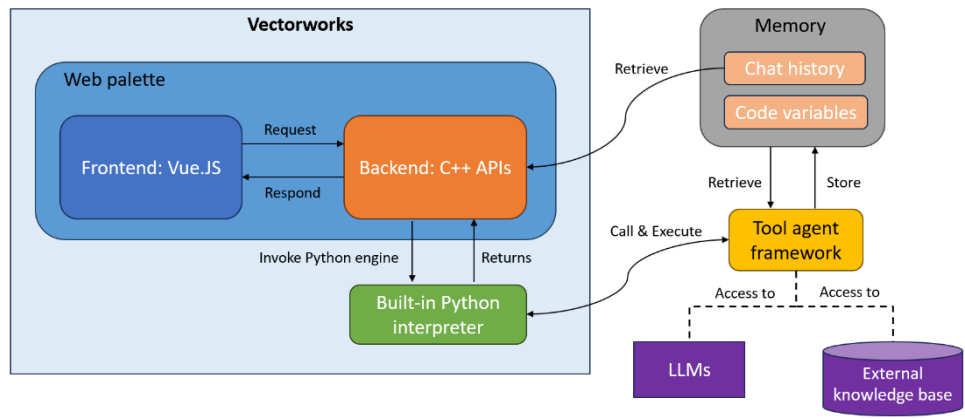

Figure 4 Software architecture of the copilot prototype in Vectorworks based on web palette plugin template[5]

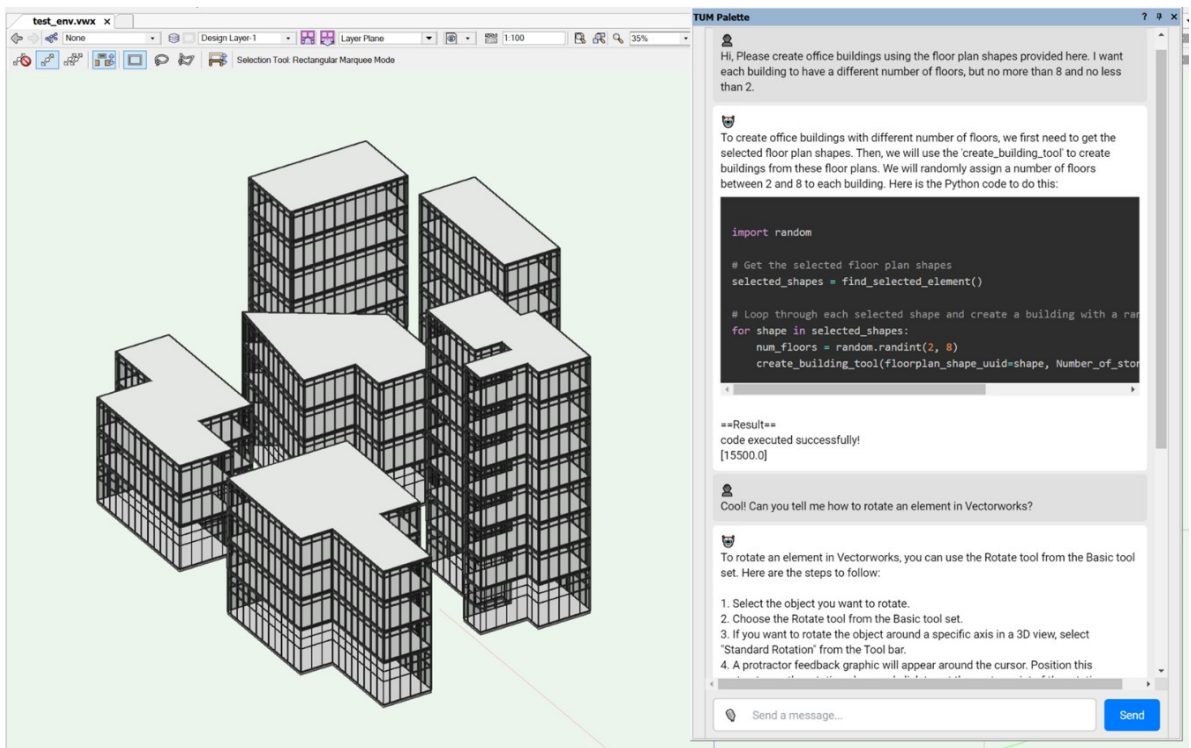

Figure 5 Seamless integration into Vectorworks. Users can interact with the copilot in the built-in chat window, giving modeling instructions or asking usage questions through voice or text. By clicking the microphone button, the backend automatically calls the Whisper model to convert the audio to text and fill the input message box.

## 4.1 Empirical evaluation

We designed several representative test prompts to empirically evaluate whether the proposed LLM agents can understand the combination of complex intent instructions to complete modeling tasks using the correct set of tools. The results are shown in Figure 6. We designed the test prompts A, B, and C as a series of consecutive dialogues, aiming to examine whether the agents can fully utilize the context information in multi-round conversations through the proposed memory module. It can be observed that GPT-4 and open-source model Mixtral can generate stable and correct results for straightforward complex instructions (A, B). Interestingly, their concepts of "north" differ - GPT-4 opts for the positive Y-axis, while Mixtral chooses the positive X-axis. However, for more abstract instructions like arranging rooms in a hotel layout (C), GPT-4 demonstrates better understanding and reasoning abilities, capable of generating the correct coordinates in code, whereas Mixtral struggles with comprehending the spatial layout. This is mainly due to the significant difference in their parameter sizes. However, it also shows that LLMs pre-trained on massive text data have a certain perceptual ability for spatial and geometric concepts. The test prompt D assesses whether the agent can interpret effective information from human intent and align it with input parameters when calling complex parametric tools. GPT-4 excels at this task, while Mixtral often encounters code errors, such as invoking non-existent or wrong tools. Interestingly, when prompted to resolve these errors, the agent can automatically correct itself and produce the right result based on the previous code and error messages the custom Python interpreter returned.

---

[5] https://github.com/VectorworksDeveloper/SDKExamples/tree/master/Examples/WebPaletteExample

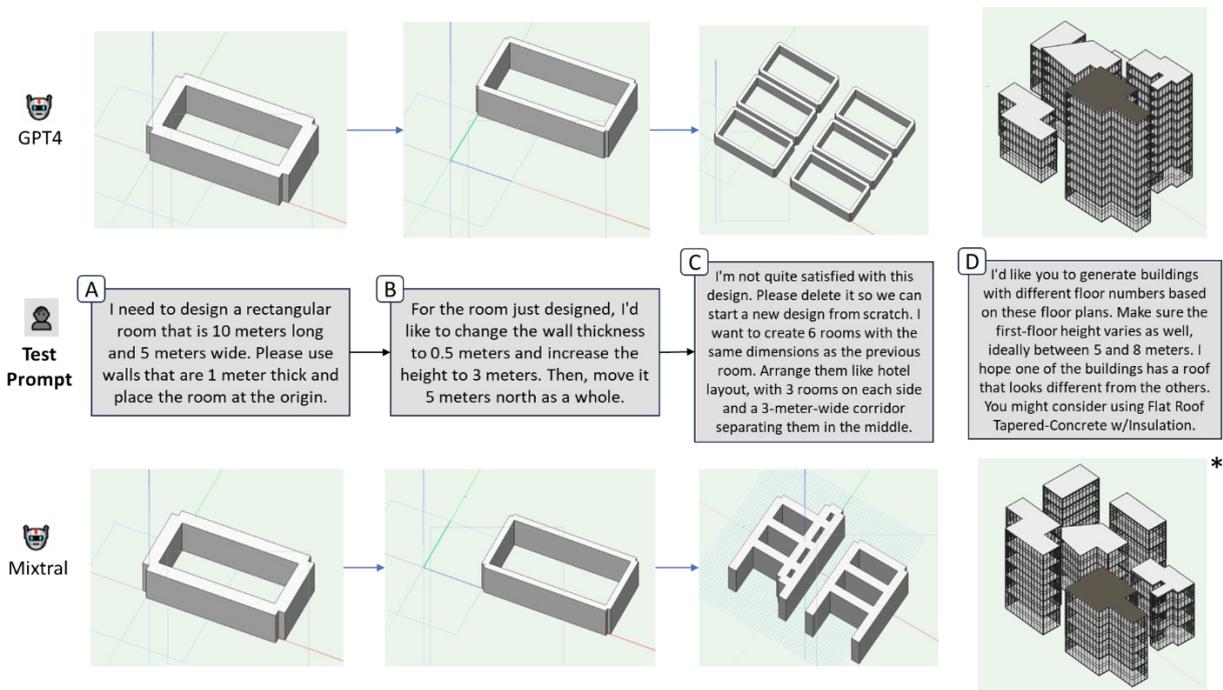

Figure 6 Test prompts and the corresponding modeling results generated by different LLM agents. The red, green and blue axes in the model represent the X, Y and Z axes, respectively. Prompts A, B, and C are sequentially related, forming a continuous dialogue, whereas prompt D is standalone. The * indicates that the agent obtained this result after revising its code based on human feedback. Overall, GPT-4 can generate more accurate and robust results compared to the latest open-source model.

We quantitatively assessed the RAG workflow designed for software usage Q&A using the RAGAs evaluation framework (Es *et al.*, 2024). We had ChatGPT pose as a user and ask 20 questions about Vectorworks, covering basic usage, advanced features, troubleshooting, etc. A synthetic validation set was created after manually correcting some hallucination issues. We comprehensively evaluated the performance of various components in our RAG pipeline using the faithfulness, context utilization, and answer relevancy metrics provided by the RAGAs framework. Faithfulness measures the consistency of the generated answers with the factual content in the given context, answer relevancy focuses on assessing how relevant the generated answers are to the given query, and context utilization calculates whether the retrieved context can be used to answer the query. The results in Table 2 demonstrate that our agent can reliably answer software usage questions based on external knowledge.

Table 2 Average evaluation metrics based on 20 usage questions

| Faithfulness | Context utilization | Answer relevancy |
|---|---|---|
| 99.5% | 100% | 96.4% |

## 5. Discussion and future works

We believe the proposed framework can be extended to a broader range of use cases if more tools are implemented for use by LLM agents. The manually defined tool functions can essentially be seen as higher-order, concise API interfaces exposed to LLMs, encapsulating specific design rules and engineering logic. This avoids the tedium of low-level API calls while ensuring the accuracy of the modeling tasks for which the tools are responsible. However, designing universal tool functions to cover different scenario needs efficiently is challenging.

In addition, the current agent invokes and combines tools based on a limited number of tool descriptions to meet user intents. Allowing it to discern and assemble the right tools from a large set could lead to unreliable call results due to hallucination issues. Therefore, for more complex design scenarios, developing a structured toolset based on certain rules could better assist agents in selection, planning, and reasoning.

Moreover, the agent currently perceives its environment and self-corrects based solely on the code execution results of the Python interpreter and human feedback. We believe providing a comprehensive building context, such as project information and component attributes, could help it perform tasks more effectively.

Finally, while the open-source LLM underperforms in highly complex tasks, its ability to be fine-tuned offers great optimization potential in our specific verticals. Additionally, deploying an instructions-tuned open-source model can better protect user data and privacy.

## 6. Conclusion

In this paper, we introduce an LLM-based agent framework that autonomously completes modeling tasks and provides suggestions for practical software usage within BIM authoring software. Our experiments employed representative complex instructions to evaluate the proposed framework, demonstrating the agent's perception of spatial and geometric concepts, ability to plan and reason based on complex prompts, utilization of external knowledge bases, and the capability to self-correct based on contextual information and human feedback during the conversation. In a case study, we developed a software prototype in Vectorworks to integrate the LLM agent as a design copilot into the user's workflow, laying a foundation for more intelligent human-machine interaction and a move towards modeling-by-chatting.

**Acknowledgment**

This work is funded by Nemetschek Group, which is gratefully acknowledged. We sincerely appreciate the data and licensing support provided by Vectorworks, Inc.